\documentclass[twocolumn,showpacs,preprintnumbers,amssymb]{revtex4}

\usepackage{graphicx}
\usepackage{epsfig}
\usepackage{dcolumn}
\usepackage{bm}
\def\gsim{\lower0.5ex\hbox{$\:\buildrel >\over\sim\:$}}
\def\lsim{\lower0.5ex\hbox{$\:\buildrel <\over\sim\:$}}

\begin{document}

\preprint{AMES-HET-03-06}   
\preprint{BNL-HET-03/19}


\title{Three heavy jet events at hadron colliders as a sensitive probe 
of the Higgs sector}

\author{David Atwood}%
\email{atwood@iastate.edu}
\affiliation{Department of Physics and Astronomy, Iowa State University, Ames,
IA 50011, USA}
\author{Shaouly Bar-Shalom}%
\email{shaouly@physics.technion.ac.il}
\affiliation{Physics Department, Technion-Institute of Technology, Haifa 32000, Israel}
\author{Gad Eilam}
\email{eilam@physics.technion.ac.il}
\affiliation{Physics Department, Technion-Institute of Technology, Haifa 32000, Israel}
\author{Amarjit Soni}%
\email{soni@bnl.gov}
\affiliation{Theory Group, Brookhaven National Laboratory, Upton, NY 11973, USA}
\date{\today}

\begin{abstract}
Assuming that a non-standard neutral Higgs with an enhanced Yukawa coupling
to a bottom quark is observed at future hadron experiments,
we propose a method for 
a better understanding of the Higgs sector. Our procedure is 
based on ``counting'' the number of events with heavy jets 
(where ``heavy'' stands for a
$c$ or $b$ jet) versus $b$ jets, in the final state of processes in which 
the Higgs is produced in association with a single high $p_T$ 
$c$ or $b$ jet. We show that an 
observed signal of the type proposed, at either the Tevatron or the LHC, 
will {\rm rule out} the popular two Higgs doublet
model of type II as well as its supersymmeric version - the Minimal 
Supersymmetric Standard Model (MSSM), and may 
provide new evidence in favor of some more exotic multi Higgs scenarios. 
As an example, we show that in a version of a two Higgs doublet model 
which naturally accounts for the large mass of the top quark,  
our signal can be easily detected 
at the LHC within that framework. 
We also find that such a signal may be 
observable at the upgraded Tevatron RunIII, 
if the neutral Higgs in this model has a mass around 100 GeV 
and $\tan\beta \gsim 50$ and if the efficiency for distinguishing
a $c$ jet from a light jet reaches the level of $\sim 50\%$.

\end{abstract}

\pacs{12.15.Ji, 12.60.Fr, 12.60.-i, 14.80.Cp}
 
\maketitle

\section{Introduction}

With the Tevatron RunII underway at Fermilab, 
the future Tevatron upgrade RunIII and the Large Hadron Collider (LHC) at CERN,
the discovery of the Higgs boson is currently 
believed to be ``around the corner''. 
These machines will be able to detect the neutral Higgs 
over its entire theoretically allowed mass range \cite{hadhiggs}.      

If the Higgs Yukawa coupling to a bottom quark 
is enhanced, as predicted for example in general multi Higgs models and 
in the supersymmetric version of the 
two Higgs Doublet Model (2HDM) - the MSSM - when 
$\tan\beta$ is large, then the neutral Higgs may be 
produced at Leading Order (LO) through 
$b \bar b$ fusion $b \bar b \to h$ \cite{bh3}. In such models,
a sizable production rate of a neutral Higgs 
in association with $b$ quark jets is also expected  
via the sub-processes $gb \to b h$ \cite{bh1,bh2} and 
$gg,~q \bar q \to b \bar b h$ \cite{bbh}.$^{[1]}$\footnotetext[1]{Unless 
stated otherwise, 
$h$ will be used to denote the neutral Higgs that drives the processes 
under consideration.} 
In particular, as was shown in \cite{bh1}, if
one demands that {\it only} one $b$ jet will be produced 
(in association with the Higgs) at high $p_T$, 
then
$gb \to b h$ becomes the LO and is therefore the dominant
Higgs-bottom associated production mechanism for 
large $\tan\beta$. For example,
$gb \to b h$ followed by $h\to b \bar b$ may already prove to be a useful    
probe of the neutral Higgs sector at the Tevatron future runs 
\cite{datta}. 

In this work we assume that indeed the Higgs has an enhanced coupling to 
bottom quarks and that it will be therefore discovered or observed 
in association with bottom quarks jets, 
via one of the above production channels.
Discovery of such a non-standard Higgs should start an exploration era of 
the various Higgs Yukawa couplings to fermions in order to decipher 
the origin and detailed properties of the Higgs sector. 
For that purpose, special observables should be devised that will enable 
the experimentalists to pin-down some specific Higgs Yukawa interactions.
In particular, one should find observables 
which are sensitive to a specific
Yukawa coupling or combination of couplings, and are therefore able 
to exclude models with an enhanced 
Higgs-bottom Yukawa vertex, or provide further evidence 
in favor of a definite Higgs scenario.

In this paper we follow this line of thought and present a method which is 
useful for an investigation of the relation between the Higgs-charm 
and Higgs-bottom Yukawa couplings. 
We focus on neutral Higgs production via a 
single high $p_T$ charm or bottom jet, $qg \to qh$, $q=c$ or $b$, 
followed by $h \to b \bar b$ or 
$h \to c \bar c$.$^{[2]}$\footnotetext[2]{We will not explicitly write 
the charge of the quarks involved. All charge-conjugate 
sub-processes are understood throughout this paper.} 
Our final state is therefore 
defined to be composed out of {\it exactly} three non-light jets, 
where a light jet means a $u,d,s$ or $g$ (gluon) jet.
If the Higgs is not the Standard Model (SM) 
one, particularly if the Higgs-bottom Yukawa is large, 
then this class of processes may play an important 
role in Higgs searches, discovery and characterization at hadron colliders. 
  
Based on this type of production mechanism, we suggest an 
observable which is particularly sensitive to new physics 
in the ratio between the Higgs-charm and the Higgs-bottom Yukawa 
couplings. We find that our signal is insensitive to the Higgs 
sectors of the type II 2HDM (2HDMII) and of the MSSM. Therefore, a positive 
signal of our observable at either the Tevatron or the LHC will rule out 
these models. 
We also consider a different type of a 2HDM, the so called  
``2HDM for the top quark'' (T2HDM), which is an interesting model 
designed to naturally accommodate the large mass of the top quark.
This model may be viewed as an effective low energy theory 
wherein the Yukawa interactions mimic some high energy dynamics 
which generates both the top mass and the weak scale. 
The two scalar doublets could be composite, as in top-color models,
and the Yukawa interactions could be the residual effect of some 
higher energy
four-Fermi operators. Interestingly, for this T2HDM our 
signal is significantly enhanced.
We thus show that if this version of a
2HDM underlies the
Higgs sector, then our signal will be easily detected (to many standard 
deviations) at the LHC and 
may also be detected at the Tevatron if $\tan\beta$ is large enough 
and if the efficiency of tagging a $c$ jet as a heavy jet 
is of ${\cal O}(50\%)$.    
   
The paper is organized as follows: in section II we define our signal. 
In section III we present the numerical setup and discuss the 
experimental background.
In section IV we evaluate our signal in three different versions 
of a 2HDM and discuss the expectations from the T2HDM, the 2HDMII 
and the MSSM. 
In section V we summarize our results. 

\section{Notation and definition of the signal}

Let $j_h$ denote a ``heavy'' $b$ or $c$-quark jet. Specifically, 
a $j_h$ jet should not be a $u$, $d$, $s$ or gluon jet 
(i.e., with the veto $j_h \neq u,d,s,g$ jet). 
Let $j_b$ be a $b$-quark jet. Clearly, every jet of type $j_b$ 
is also a $j_h$. Likewise, let us denote by $j_c$ a $c$-quark jet.  

We thus define the cross-sections

\begin{eqnarray}
\sigma_{j_hj_hj_h} &\equiv& \sigma(pp~{\rm or}~p \bar p\to j_hj_hj_h +X)~,\\
\sigma_{j_bj_bj_b} &\equiv& \sigma(pp~{\rm or}~p \bar p\to j_bj_bj_b +X)~,
\end{eqnarray}

\noindent and the ratio

\begin{equation}
{\cal R} \equiv 
\frac{\sigma_{j_hj_hj_h}}{\sigma_{j_bj_bj_b}} \label{calR}~.
\end{equation}

\noindent Note that, since $j_h$ is either a $j_b$ or a $j_c$, 
it follows that:  

\begin{equation}
\sigma_{j_hj_hj_h} \equiv \sigma_{j_bj_bj_b}+\sigma_{j_cj_bj_b}
+\sigma_{j_bj_cj_c}+\sigma_{j_cj_cj_c} ~.
\end{equation}

\noindent From the above definitions it is evident that: 

\begin{equation}
{\cal R}-1 \propto \frac{Y_c^2}{Y_b^2} + 
{\cal O} \left( \frac{Y_c^4}{Y_b^4} \right) ~.
\end{equation}

\noindent where $Y_c$ and $Y_b$ are the Higgs-charm and Higgs-bottom 
Yukawa couplings, respectively.

In reality, one has to include non-ideal efficiencies for jet 
identification. Therefore, we define the quantity $\epsilon_j^k$ to be 
the efficiency for identifying a $j$ jet as a $k$ jet. 
Thus, for example,
$\epsilon_l^b$ or $\epsilon_l^h$ are the efficiencies 
for misidentifying a light jet ($j_l$) as 
a $b$ jet ($j_b$) or as a heavy jet ($j_h$), respectively. 
Using these efficiency factors we define the experimental observable as:

\begin{equation}
R^M \equiv 
\frac{\bar\sigma^M_{j_hj_hj_h}}{\bar\sigma^M_{j_bj_bj_b}} \label{eq6} ~,
\end{equation}

\noindent where $R=\bar\sigma_{j_hj_hj_h}/\bar\sigma_{j_bj_bj_b}$ 
with $\bar\sigma_{j_hj_hj_h}$, 
$\bar\sigma_{j_bj_bj_b}$ being the 
effective signals for cross-sections that will  actually be measured in 
the experiment  
and $R^M$ is the value of $R$ calculated 
within a specific Higgs model M via 
the sub-processes $g q \to q h \to q q q$, $q=c$ or $b$. 
Thus, in the limit of an ideal detector where $\epsilon^k_j=\delta^k_j$, 
$R^M = {\cal R}^M$ (where ${\cal R}^M$ means the ``ideal'' ratio 
defined in (\ref{calR}) calculated within model M).  
In particular, these cross-sections include all tagging efficiencies as
follows: 

\begin{eqnarray}
\bar\sigma^M_{j_bj_bj_b}&=&\sum_{j_1,j_2,j_3} 
\epsilon_{j_1}^b\epsilon_{j_2}^b\epsilon_{j_3}^b \times 
\sigma^M_{j_1j_2j_3} ~,\\
\bar\sigma^M_{j_hj_hj_h}&=&\sum_{j_1,j_2,j_3} 
\epsilon_{j_1}^h\epsilon_{j_2}^h\epsilon_{j_3}^h \times 
\sigma^M_{j_1j_2j_3} \label{eq8} ~.
\end{eqnarray}

\noindent In the limit $(Y_{c}/Y_{b})^2 \to 0$,  
one obtains 

\begin{equation}
R^M \to R^0 \equiv \frac{\bar\sigma^M_{j_hj_hj_h}}{\bar\sigma^M_{j_bj_bj_b}}
\left|_{\left(\frac{Y_c}{Y_b}\right)^2\to 0} \right. \to
\frac{(\epsilon_b^h)^3\sigma^M_{j_bj_bj_b}}{(\epsilon_b^b)^3 
\sigma^M_{j_bj_bj_b}} = \left(\frac{\epsilon_b^h}{\epsilon_b^b}\right)^3 ~.
\end{equation}
 
\noindent Therefore, due to the non-ideal 
efficiencies (in particular if $\epsilon_b^h \neq \epsilon_b^b$), 
we have $R^0 \neq 1$  
even if the Higgs Yukawa coupling
to the charm quark is negligible.
As will be shown below, this limit is applicable 
to a neutral Higgs either of MSSM 
or of 2HDMII origin, 
if $\tan\beta$ is large. That is, for large $\tan\beta$ 

\begin{equation}
R^{MSSM},~R^{2HDMII} \to R^0 = 
\left(\frac{\epsilon_b^h}{\epsilon_b^b}\right)^3 \label{susyrel}~.
\end{equation} 
  
\noindent A deviation from $R^M=R^0$, indicating a high rate 
for $h \to c \bar c$, signals a specific type 
of new physics in the Higgs sector, e.g., beyond the MSSM or the 2HDMII 
Higgs scenario. Such a deviation is parameterized by:

\begin{equation}
\delta R^M = R^M-R^0 \propto \left(\frac{Y_c}{Y_b}\right)^2 +
{\cal O} \left( \frac{Y_c^4}{Y_b^4} \right) \label{delRth} ~.
\end{equation}

\noindent The significance with which $\delta R^M$ can be measured depends   
on the experimental error $\delta R^M_{\rm exp}$:

\begin{equation}
\delta R^M_{\rm exp}=R^M \sqrt{ \frac{(\Delta
N_{j_hj_hj_h})^2}{(N_{j_hj_hj_h}^M)^2}+\frac{(\Delta
N_{j_bj_bj_b})^2}{(N_{j_bj_bj_b}^M)^2} } \label{delRMexp}~,
\end{equation}

\noindent where $N_{j_hj_hj_h}^M=\bar\sigma^M_{j_hj_hj_h} \times L$ 
and similarly for $N_{j_bj_bj_b}^M$, with
$L$ being the integrated luminosity at the given collider.
The errors in the measurements of $N_{j_hj_hj_h}^M$ and 
$N_{j_bj_bj_b}^M$ are (statistical only): 
$\Delta N_{j_hj_hj_h} = \sqrt{N_{j_hj_hj_h}}$ and
$\Delta N_{j_bj_bj_b} = \sqrt{N_{j_bj_bj_b}}$, 
where $N_{j_hj_hj_h}$ and $N_{j_bj_bj_b}$ 
(i.e., without the superscript M) are 
the total number of events dominated by the background processes. 

Eq.(\ref{delRMexp}) can be simplified to:

\begin{equation}
\delta R^M_{\rm exp} = 
\frac{\sqrt{N_{j_hj_hj_h}+(R^M)^2 N_{j_bj_bj_b}}}{N_{j_bj_bj_b}^M}
\label{delRexp}~,
\end{equation} 

\noindent therefore, since $N_{j_bj_bj_b}^M \propto Y_b^2$, 
it follows that $\delta R^M_{\rm exp} \propto 1/Y_b^2$. In particular, 
if $Y_b \propto \tan\beta$, then
$\delta R^M_{\rm exp} \propto 1/\tan^2\beta$.   
    
The condition, $\delta R^M > \delta R^M_{\rm exp}$ 
signals that new physics in the ratio between the $hcc$ and $hbb$ Yukawa
couplings can be seen, with a statistical significance of

\begin{equation}
N_{SD}=\frac{\delta R^M}{\delta R_{\rm exp}^M} \label{NSD}~.
\end{equation}

\section{Numerical setup and background \label{sec3}}

Before presenting our results let us describe the numerical setup: 
\begin{itemize}

\item All signal and background cross-sections are calculated 
at leading order (tree-level), 
using the COMPHEP package \cite{comphep} with the CTEQ5L \cite{cteq5l} parton 
distribution functions.

\item Throughout our entire analysis to follow, 
the following set of cuts are employed in {\it both} 
signal and background cross-sections:
\begin{enumerate}
\item For both the Tevatron and the LHC, in order for 
the jets to be within the tagging region of the silicon vertex detector, 
we require all three jets to have the following minimum transverse momentum 
($p_T$) and rapidity ($\eta$) coverages:\\ 
Tevatron: $p_T(j)>15~{\rm GeV}~,~|\eta(j)|<2$,\\
LHC:      $~~~~~~p_T(j)>30~{\rm GeV}~,~|\eta(j)|<2.5$,\\
\noindent where $j$ is any of the three jets in the final state.

\item The signal cross-sections always have one pair of jets (coming from the 
Higgs decay) that should reconstruct 
the Higgs mass. Thus, in order to improve the signal to background ratio, 
we require for both the signal and background cross-sections 
that {\it only one} jet pair (out of the 
three possible jet pairs in the final state), will have an invariant mass
within $m_h \pm 0.05 m_h$, i.e., within $\pm 5\%$ of $m_h$
around $m_h$. That is, we reject events if the invariant masses 
of {\it more than} one 
$j_ij_k$ pair are within 
$m_h - 0.05 m_h <M_{j_i j_k}<m_h + 0.05 m_h$. This acceptance cut
is found to significantly improve the signal to background ratio.

\item We impose a common lower cut on the invariant masses of 
all three jet pairs: $M_{j_i j_k} > m_h/2$.   
This additional cut further improves the signal to background ratio.

\item We require the final state partons to be separated by a cone 
angle of $\Delta R > 0.7$, where here $\Delta R$ stands for 
$\sqrt{\Delta\eta^2+\Delta\varphi^2}$.
\end{enumerate}

\item We use the $\overline{MS}$ 
running Yukawa couplings $\bar Y_b(\mu_R) \propto (\bar m_b(\mu_R)/v)$, 
$\bar Y_c(\mu_R) \propto (\bar m_c(\mu_R)/v)$, 
where $\bar m_b(\mu_R)$ and $\bar m_c(\mu_R)$ are 
$\overline{MS}$ running masses. We take a 
renormalization scale of $\mu_R=m_h$ as our 
central value.
In particular, $\bar m_b(m_h)$ and $\bar m_c(m_h)$ are calculated 
via the next-to-leading-order heavy quark $\overline{MS}$ running 
mass equation \cite{MSrun}, 
using $\bar m_b(\bar m_b)$ and $\bar m_c(\bar m_c)$ as the initial 
conditions. This brings up some uncertainty corresponding to the 
allowed range of $\bar m_b(\bar m_b)$ and $\bar m_c(\bar m_c)$: 
$4 ~{\rm GeV}< \bar m_b(\bar m_b) < 4.5 ~{\rm GeV}$ and 
$1 ~{\rm GeV}< \bar m_c(\bar m_c) < 1.4 ~{\rm GeV}$, see \cite{PDG}.
In what follows we use $\bar m_b(\bar m_b)=4.26$ GeV 
and $\bar m_c(\bar m_c)=1.26$ GeV as our central values \cite{PDG}, which 
give the $\bar m_b(m_h)$ and $\bar m_c(m_h)$ masses 
listed in Table \ref{runmass}.

\begin{table}[htb]
\begin{center}
\caption[first entry]{Running charm and bottom 
quark $\overline{MS}$ (NLO) masses 
at $\mu_R=m_h$ for $m_h=100,~120$ and $140$ GeV, with
the initial conditions 
$\bar m_b(\bar m_b)=4.26$ GeV and $\bar m_c(\bar m_c)=1.26$ GeV.
\bigskip
\protect\label{runmass}}
\begin{tabular}{|c||c|c|c|} 
\multicolumn{4}{|c|}{running charm and bottom $\overline{MS}$ masses} \\
\multicolumn{4}{|c|}{$\bar m_b(\bar m_b)=4.26$ GeV, $\bar m_c(\bar m_c)=1.26$ GeV} \\ \hline \hline  
$~\Downarrow$[GeV]$\Rightarrow~$& $~\mu_R=100~$ & $~\mu_R=120~$  &$~\mu_R=140~$\\ \hline 
$\bar m_b(\mu_R)$ & $2.94$    & $2.89$    & $2.86$  \\ 
$\bar m_c(\mu_R)$ & $0.61$  & $0.6$   & $0.59$ \\ \cline{1-4}
\end{tabular}
\end{center}
\end{table}
 
At some instances we will refer to the ``maximal signal'' possible. 
This means that we are evaluating the signal cross-sections with 
the initial conditions $\bar m_b(\bar m_b)=4.5$ GeV 
and $\bar m_c(\bar m_c)=1.4$ GeV, i.e., at their upper allowed values, 
which at $\mu_R=100$ GeV gives: $\bar m_b(100~{\rm GeV})=3.1$ GeV and 
$\bar m_c(100~{\rm GeV})=0.68$ GeV.    

\item Unless stated otherwise, the parton distribution 
functions are evaluated with a factorization 
scale of $\mu_F=m_h$. Thus, we use a common 
factorization and renormalization scale $\mu=\mu_F=\mu_R$ and 
take $\mu=m_h$ as our nominal value. The uncertainty of our signal 
cross-sections, obtained by varying the factorization scale $\mu_F$ about 
its central value ($\mu_F=m_h$) within the range $m_h/4 < \mu_F < 2m_h$, 
will also be investigated.        

\item We do not consider the next-to-leading-order (NLO) contributions 
to $\sigma(qg \to qh)$.          
As was shown in \cite{bh1}, the NLO contribution 
to $\sigma(bg \to bh)$
can amount to an effective $K$-factor of about 
$1.5$ at the Tevatron and about $1.2$ at the LHC. Assuming that a similar 
$K$-factor applies also to $\sigma(cg \to ch)$, our signal $R^{M}$ 
is essentially insensitive to the NLO corrections since it involves ratios 
of these cross-sections. 
Moreover, $\delta R_{\rm exp}^M$ is proportional to $\sqrt{K^B}/K^S$, where 
$K^B$ and $K^S$ are the $K$-factors for the background and the 
signal three jets events, 
respectively. Thus, at NLO, the statistical significance of our signal
($N_{SD}=\delta R^M/\delta R_{\rm exp}^M$) 
should be about a factor 
of $K^S/\sqrt{K^B}$ larger or smaller than our tree-level estimate, 
depending on  
the size of $K^B$ and $K^S$. For $K^B \sim K^S > 1$, our tree-level 
prediction for $N_{SD}$ is on the conservative side.

\item Since we require our signal to be 
composed out of events with {\it exactly} three jets in the final state
(i.e., 3 heavy jets, $j_hj_hj_h$), 
we do not consider 
processes with 4 heavy jets or 4 $b$ jets in the final state, e.g., 
$gg,~q\bar q \to b\bar bh \to b \bar b b \bar b$. 
These type of sub-processes constitute a 
partial source of the $K$ factor for the 3 heavy jets signal 
cross-sections \cite{bh1}. 
Moreover, as was further demonstrated in \cite{bh1} for 
the case of Higgs-bottom associated production at hadron colliders, 
even if we had relaxed our demand for 
{\it exactly} 3 jets in the final state to allow for 3 or 4 jets 
with 3 or 4 jet tags, 
the contribution of events with 4 jets to $R^M$ would 
still constitute no more than $10\%$ of those with exactly 3 jets 
in the final state. 

\end{itemize}

Let us now discuss the background cross-sections. 
The irreducible background for the various
three heavy jet final states, $j_hj_hj_h$ with $h=c$ or $b$, 
is dominated by the QCD sub-processes
$b g \to b g \to bbb,~bcc$ and $c g \to c g \to ccc,~cbb$. Other 
processes like  
$Z$ production followed by its decay,        
$b g \to b Z \to bbb,~bcc$ and $c g \to c Z \to ccc,~cbb$, 
and QED also contribute to the background. 
In what follows, this irreducible background
is calculated using the COMPHEP package \cite{comphep}, where   
all possible tree-level
diagrams (QCD, Electroweak and QED) for the 
$j_hj_hj_h$ final state are included. 
In addition, due to the non-ideal efficiencies, e.g., the non-zero probability
of misidentifying a light jet for a heavy jet, 
there is a reducible background for the three heavy jet 
final states coming from sub-processes in which one, two or all three 
jets are light. Since one expects that 
$\epsilon_l^h,~\epsilon_l^b \ll \epsilon_b^h,~\epsilon_b^b,~\epsilon_c^h,~\epsilon_c^b$ (see e.g., \cite{epsnumbers}), the
reducible background 
cross-sections, when multiplied by $\epsilon_l^h$ or $\epsilon_l^b$ 
are dominated 
by processes in which only one jet out of the
three is light. Moreover, we find that, at both the Tevatron and the LHC, 
the one-light-jet reducible background is by far controlled by the
$gg \to g bb$ and $gg \to gcc$ sub-processes, which will be therefore 
included in our background estimation. For example, the contribution
of $gg \to g bb$ to the $j_hj_hj_h$ final state is: 
$\bar\sigma_{j_hj_hj_h}^B=\sigma_{gg \to g bb} \times 
\epsilon_l^h (\epsilon_b^h)^2$ (the background signals will always be 
denoted by the superscript $B$). 

In Tables \ref{tabbcgTEV} and \ref{tabbcgLHC} we list the 
background cross-sections at the Tevatron and at the LHC, respectively, 
where all cross-sections are calculated with the set of cuts 1-4, 
described earlier. The background cross-sections 
are calculated with a factorization scale set to $\mu_F=\sqrt{\hat s}$, where 
$\hat s$ is the square of the c.m. energy of the hard cross-sections.
  
\begin{table}[htb]
\begin{center}
\caption[first entry]{Background cross-sections in [fb], 
for the Tevatron with a c.m. of $\sqrt{s}=2$ TeV. All cross-sections
are calculated with the following kinematical cuts (some defined 
as a function of $m_h$):  
(1) $p_T(j)>15$ GeV, (2) $|\eta_j|<2$, where $j$ is 
any (heavy and light) of the three jets 
in the final state, (3) the acceptance cut that 
the invariant mass of only one jet pair $M_{j_i j_k}$, 
$i\neq k$, out of the three possible pairs of 
jets in the final state is within the mass range 
$(m_h - 0.05 m_h) <M_{j_i j_k}<(m_h + 0.05 m_h)$, (4) 
$M_{j_i j_k}> m_h/2$, for all three jet pairs, and (5) any two 
partons in the final state are separated by a cone of $\Delta R >0.7$.  
Numbers are rounded to the $1\%$ accuracy as was  
obtained from the COMPHEP numerical sessions.
\bigskip
\protect\label{tabbcgTEV}}
\begin{tabular}{|c||c|c|c|}
\multicolumn{4}{|c|}{Background cross-sections at the Tevatron} \\ \hline
& \multicolumn{3}{c|}{$m_h$ used for the kinematical cuts} \\ \cline{2-4}
[fb]$\Downarrow$ / [GeV]$\Rightarrow$& $m_h=100$ & $m_h=120$  &$m_h=140$\\ \hline \hline
$\sigma^{B}(gb \to bbb~{\rm or}~bcc)$ & $4.4\cdot10^3$ & $1.8\cdot10^3$ & $9.5\cdot10^2$ \\ 
$\sigma^{B}(gc \to cbb~{\rm or}~ccc)$ & $7.2\cdot10^3$ & $3.0\cdot10^3$ & $16.3\cdot10^2$ \\ 
$\sigma^{B}(gg \to gbb~{\rm or}~gcc)$ & $230\cdot10^3$ & $94\cdot10^3$ & $41\cdot10^3$  \\ \cline{1-4}
\end{tabular}
\end{center}
\end{table}

\begin{table}[htb]
\begin{center}
\caption[first entry]{Background cross-sections in [fb], 
for the LHC with a c.m. of $\sqrt{s}=14$ TeV. All cross-sections
are calculated with $p_T(j)>30$ GeV, $|\eta_j|<2.5$ and the additional 
kinematical cuts 3-5 
as in Table \ref{tabbcgTEV} (see caption to Table \ref{tabbcgTEV}).
\bigskip
\protect\label{tabbcgLHC}}
\begin{tabular}{|r||r|r|r|}
\multicolumn{4}{|c|}{Background cross-sections at the LHC} \\ \hline
& \multicolumn{3}{c|}{$m_h$ used for the kinematical cuts} \\ \cline{2-4}
[fb]$\Downarrow$ / [GeV]$\Rightarrow$& $m_h=100$ & $m_h=120$  &$m_h=140$\\ \hline \hline
$\sigma^{B}(gb \to bbb~{\rm or}~bcc)$ & $133\cdot10^3$ & $111\cdot10^3$ & $84\cdot10^3$ \\ 
$\sigma^{B}(gc \to cbb~{\rm or}~ccc)$ & $190\cdot10^3$ & $159\cdot10^3$ & $120\cdot10^3$ \\
$\sigma^{B}(gg \to gbb~{\rm or}~gcc)$ & $5680\cdot10^3$ & $4720\cdot10^3$ & $3540\cdot10^3$  \\ \cline{1-4}
\end{tabular}
\end{center}
\end{table}
 
\section{Expectations from Two Higgs doublet models}  

The signal cross-sections depend on the Higgs couplings and therefore 
on the underlying Higgs model. In what follows we will 
investigate the sensitivity of our signal $R^M$ to various versions 
of a 2HDM. 

The most general 2HDM Yukawa term is given by

\begin{eqnarray} 
{\cal L}_Y &=& - \sum_{i,j}  \bar Q_L^i 
\left[ \left(U_{ij}^1 \tilde\Phi_1  +U_{ij}^2 \tilde\Phi_2 \right) u^j_R 
\right. \nonumber\\
&& \left. + 
\left(D_{ij}^1 \Phi_1  +D_{ij}^2 \Phi_2 \right) d^j_R \right] \label{yukawa1}~,
\end{eqnarray}

\noindent where $Q_L$ is the SU(2) left-handed quark doublet,
$u_R$ and $d_R$ are the right-handed up and down quark SU(2) singlets, 
respectively, and   
$\tilde\Phi_{1,2}=i \sigma_2 \Phi_{1,2}^*$. Also, $U^1,U^2,D^1,D^2$ 
are general Yukawa $3 \times 3$ matrices in flavor space.
The different types of 2HDM's are then categorized according to 
the different choices of the Yukawa matrices $U^1,U^2,D^1,D^2$. 

\subsection{The 2HDM for the top quark - T2HDM}
    
As an example of a Higgs sector that can give rise to an observable
effect via $R^M$, we consider first the so called ``2HDM for the top'', 
in which the top quark receives a special 
status \cite{T2HDM}. We will denote this model by T2HDM.
In the T2HDM one defines \cite{T2HDM}

\begin{eqnarray}
U^1_{ij} &\to& G_{ij} \times (\delta_{j1}+\delta_{j2}) ~,\nonumber \\
U^2_{ij} &\to& G_{ij} \times \delta_{j3} \label{yukawa2} ~,\\
D^2_{ij} &\to& 0 \nonumber ~,
\end{eqnarray}

\noindent where $G$ is an unknown Yukawa $3 \times 3$ matrix 
in quark flavor space. The large mass of the top is, thus,  
naturally accommodated in the T2HDM by coupling the second Higgs doublet 
($\Phi_2$), which has a much larger   
Vacuum Expectation Value (VEV), only to the top quark. This model
has therefore 
large $\tan\beta = v_2/v_1$ by construction. That is, 
$v_2$ is the VEV responsible for the top quark mass, while $v_1$ generates 
the masses of all the other quarks.

Like any other 2HDM, the Higgs spectrum of the T2HDM is composed out of
three neutral Higgs; two CP-even scalars and one pseudoscalar, and  
a charged Higgs. Choosing the basis $\alpha=\beta$, 
where $\alpha$ is the usual mixing angle between the two CP-even Higgs 
particles (see e.g., 
\cite{HHG}), the 
neutral Higgs spectrum of the T2HDM is rotated such that it contains 
a SM-like neutral Higgs, $H_{SM}$, 
(with the SM couplings to fermions), a CP-even Higgs, $h$,  
and a pseudoscalar (CP-odd Higgs), $A$.     
The Yukawa couplings of $h$ and $A$ to down and up-type quarks in the 
T2HDM are:

\begin{eqnarray}
hd \bar d&:&~Y^{SM}_d \times \tan\beta ~, \nonumber \\ 
hu \bar u&:&~Y^{SM}_u \times \tan\beta ~;~u=u~{\rm or}~c ~, \nonumber \\
ht \bar t&:&~Y^{SM}_t \times \cot\beta ~, \nonumber \\
Ad \bar d&:&~Y^{SM}_d \times \tan\beta \gamma_5 ~, \nonumber \\
Au \bar u&:&~Y^{SM}_u \times \tan\beta \gamma_5 ~;~u=u~{\rm or}~c ~, \nonumber \\
At \bar t&:&~Y^{SM}_t \times \cot\beta \gamma_5 \label{yuktop2HDM} ~.
\end{eqnarray}

\noindent where $d=d~,s$ or $b$ and $Y^{SM}_f = m_f/v$, 
$v=\sqrt{v_1^2+v_2^2}=246$ GeV.

We therefore see that while the Yukawa couplings of $h$ and $A$ to 
down quarks in the T2HDM have the same pattern 
as in the 2HDMII and the MSSM, i.e., 
$Y_d \propto Y_d^{SM} \times \tan\beta$, their 
Yukawa couplings to a $u$ and a $c$ quark exhibit a different behavior, 
due to the special structure of the T2HDM's 
Yukawa matrices in (\ref{yukawa2}). In particular, $Y_c$ and $Y_u$ 
are enhanced in the T2HDM by a factor of $\tan^2\beta$ 
as compared to their values in the 2HDMII or 
in the MSSM, in which $Y_u \propto Y_u^{SM} \times \cot\beta$ for 
$u=u,~c$ or $t$. Notice also that $Y_t$ (in the T2HDM)
remains the same as in the 2HDMII and the MSSM.

Therefore, as will be shown below, 
since $\delta R^M \propto (Y_c/Y_b)^2$ [see (\ref{delRth})], 
we can expect to have  
the following relations: $\delta R^{T2HDM} \sim  \delta R^{SM} \propto 
(m_c/m_b)^2$ and $\delta R^{2HDMII},~\delta R^{MSSM} \ll 
\delta R^{T2HDM}$, when $\tan\beta$ is large. Note that in spite of 
the fact that $\delta R^{SM} \sim \delta R^{T2HDM}$, 
due to the small production rate of a SM (or a SM-like) Higgs via the 
gluon-$b$ and gluon-$c$ fusion channels, $\delta R^{SM}$ is 
overwhelmed by the very large background (i.e., $\delta R_{\rm exp}^{SM}$) 
and is thus unobservable.

In Tables \ref{tabsigTEV} and \ref{tabsigLHC} we give  
the signal cross-sections at the Tevatron and at the LHC, respectively, 
for the T2HDM with $\tan\beta=30$ and evaluated 
at $\mu=m_h$ as the central value. To appreciate the 
uncertainties corresponding to 
variations of the factorization scale ($\mu_F$), 
we also give in Tables \ref{tabsigTEV} and \ref{tabsigLHC}
the signal cross-sections evaluated with
$\mu_F=m_h/4$ (upper uncertainty) and with $\mu_F=2m_h$ (lower uncertainty), 
while keeping a fixed renormalization scale $\mu_R=m_h$.
The effect of $\mu_F$ on the statistical significance of the signal
will be discussed below.

All cross-sections in Tables \ref{tabsigTEV} and \ref{tabsigLHC} 
are calculated with the set of cuts, as described in the 
previous section.$^{[3]}$\footnotetext[3]{The relevant Lagrangian pieces 
of the T2HDM were implemented 
into COMPHEP. Also, the Higgs width was explicitly 
calculated within the T2HDM via COMPHEP.}
To a good approximation, the signal cross-sections 
given in Tables \ref{tabsigTEV} and \ref{tabsigLHC} are  
essentially insensitive to whether the Higgs produced is 
$h$ or the pseudoscalar $A$. The difference between the two is 
the $1$ versus the $\gamma_5$ couplings to fermions, which amounts to 
sign differences in the sub-leading terms involving $m_b$ or $m_c$.

\begin{table}[htb]
\begin{center}
\caption[first entry]{Signal cross-sections in [fb], for the T2HDM with $\tan\beta=30$, 
at the Tevatron with a c.m. of $\sqrt{s}=2$ TeV. 
All cuts are the same as for the background cross-sections at the Tevatron 
(see caption to Table \ref{tabbcgTEV}). 
The upper and lower uncertainties correspond 
to $\mu_F=m_h/4$ and $\mu_F=2m_h$, respectively.

\bigskip
\protect\label{tabsigTEV}}
\begin{tabular}{|c||c|c|c|} 
\multicolumn{4}{|c|}{Signal cross-sections for the T2HDM with $\tan\beta=30$} \\ \hline
& \multicolumn{3}{c|}{Tevatron, $\sqrt{s}=2$ TeV} \\ \cline{2-4}
[fb]$\Downarrow$ / [GeV]$\Rightarrow$& $m_h=100$ & $m_h=120$  &$m_h=140$\\ \hline \hline
$\sigma^{T2HDM}(gb \to bbb)$ & $970^{+9\%}_{-9\%}$ & $385^{+15\%}_{-10\%}$ & $166^{+20\%}_{-12\%}$ \\ 
&&&\\ 
$\sigma^{T2HDM}(gb \to bcc)$ & $42^{+9\%}_{-9\%}$  & $16.6^{+15\%}_{-10\%}$   & $7^{+20\%}_{-12\%}$ \\ 
&&&\\
$\sigma^{T2HDM}(gc \to cbb)$ & $72^{+35\%}_{-14\%}$  & $28.4^{+39\%}_{-15\%}$  & $12.1^{+42\%}_{-15\%}$ \\ 
&&&\\ 
$\sigma^{T2HDM}(gc \to ccc)$ & $3.1^{+35\%}_{-14\%}$ & $1.23^{+39\%}_{-15\%}$  & $0.52^{+42\%}_{-15\%}$ \\ \cline{1-4}
\end{tabular}
\end{center}
\end{table}

\begin{table}[htb]
\begin{center}
\caption[first entry]{Signal cross-sections in [fb], for the T2HDM 
with $\tan\beta=30$, 
at the LHC with a c.m. of $\sqrt{s}=14$ TeV. 
All cuts are the same as for the background cross-sections at the LHC. 
See also captions to Table \ref{tabbcgLHC} and \ref{tabsigTEV}.

\bigskip
\protect\label{tabsigLHC}}
\begin{tabular}{|c||c|c|c|} 
\multicolumn{4}{|c|}{Signal cross-sections for the T2HDM with $\tan\beta=30$} \\ \hline
& \multicolumn{3}{c|}{LHC, $\sqrt{s}=14$ TeV} \\ \cline{2-4}
[fb]$\Downarrow$ / [GeV]$\Rightarrow$& $m_h=100$ & $m_h=120$  &$m_h=140$\\ \hline \hline
$\sigma^{T2HDM}(gb \to bbb)$ & $46000^{-18\%}_{+1\%}$ & $32400^{-12\%}_{+2\%}$ & $21900^{-9\%}_{-1\%}$ \\ 
&&&\\ 
$\sigma^{T2HDM}(gb \to bcc)$ & $2000^{-18\%}_{+1\%}$  & $1400^{-12\%}_{+2\%}$   & $940^{-9\%}_{-1\%}$ \\ 
&&&\\ 
$\sigma^{T2HDM}(gc \to cbb)$ & $3050^{+0\%}_{-4\%}$  & $2120^{+5\%}_{-3\%}$   & $1400^{+7\%}_{-5\%}$ \\ 
&&&\\
$\sigma^{T2HDM}(gc \to ccc)$ & $131^{+0\%}_{-4\%}$   & $92^{+5\%}_{-3\%}$    & $60^{+7\%}_{-5\%}$  \\ \cline{1-4}
\end{tabular}
\end{center}
\end{table}

\begin{figure*}[htb]
\psfull
 \begin{center}
  \leavevmode
  \epsfig{file=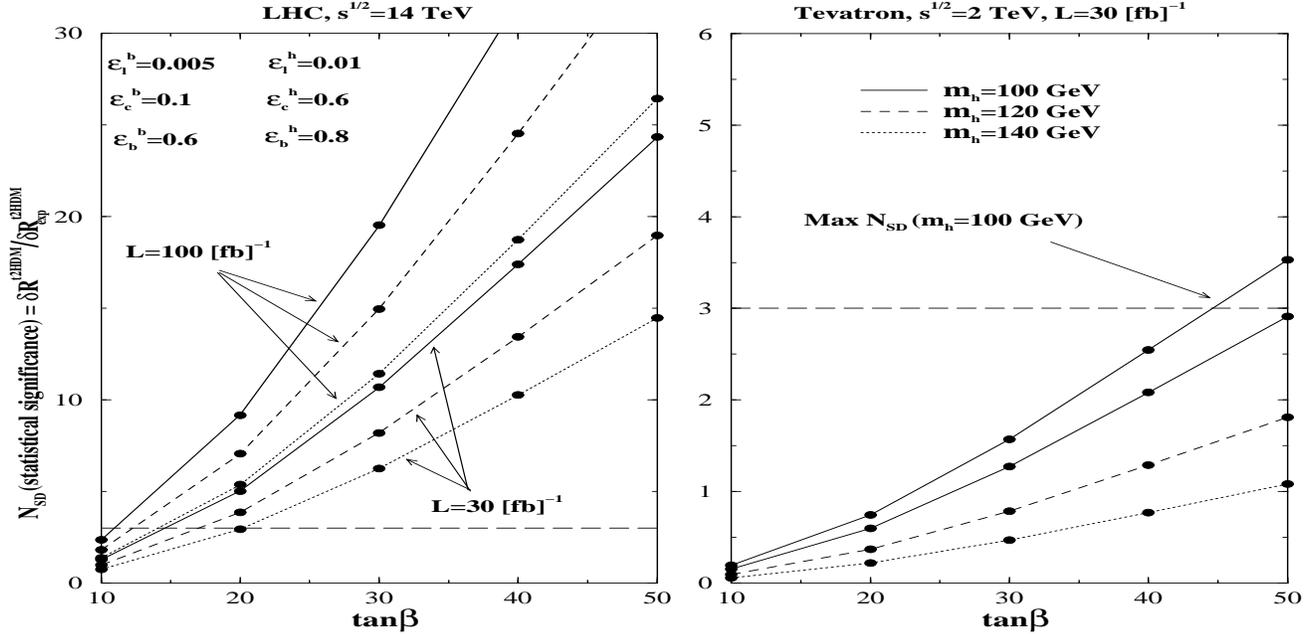,height=17cm,width=9cm,bbllx=0cm,bblly=2cm,bburx=20cm,bbury=25cm,angle=270}
 \end{center}
\caption{The statistical significance of the signal in the T2HDM 
($N_{SD}=\delta R^{T2HDM}/\delta R_{exp}^{T2HDM}$) 
as a function of $\tan\beta$ and for $m_h=100,~120$ and $140$ GeV,
at the LHC with integrated luminosity of 
30 fb$^{-1}$ and 100 fb$^{-1}$ (left plot) and at the Tevatron 
with integrated luminosity of 30 fb$^{-1}$ (right plot). The upper 
solid curve on the right plot is for the case of maximal 
charm and bottom $\overline{MS}$ quark masses at $\mu_R=m_h=100$ GeV (see 
discussion in section \ref{sec3}). 
All cross-sections were calculated for 
$\mu=\mu_F=\mu_R=m_h$ and with the set of cuts 1-4 as 
outlined in section \ref{sec3}.}
\label{fig1tb}
\end{figure*}

\begin{figure*}[htb]
\psfull
 \begin{center}
  \leavevmode
  \epsfig{file=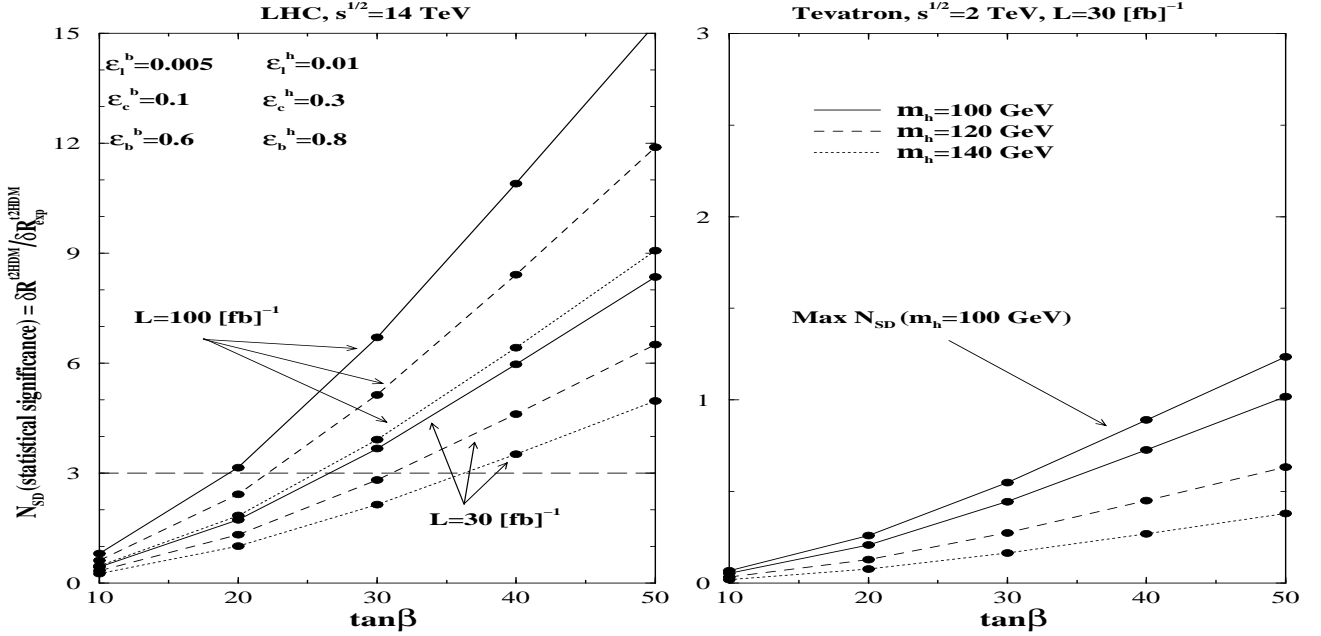,height=17cm,width=9cm,bbllx=0cm,bblly=2cm,bburx=20cm,bbury=25cm,angle=270}
 \end{center}
\caption{Same as Fig.~\ref{fig1tb}, for $\epsilon_c^h=0.3$.}
\label{fig2tb}
\end{figure*}

\begin{figure*}[htb]
\psfull
 \begin{center}
  \leavevmode
  \epsfig{file=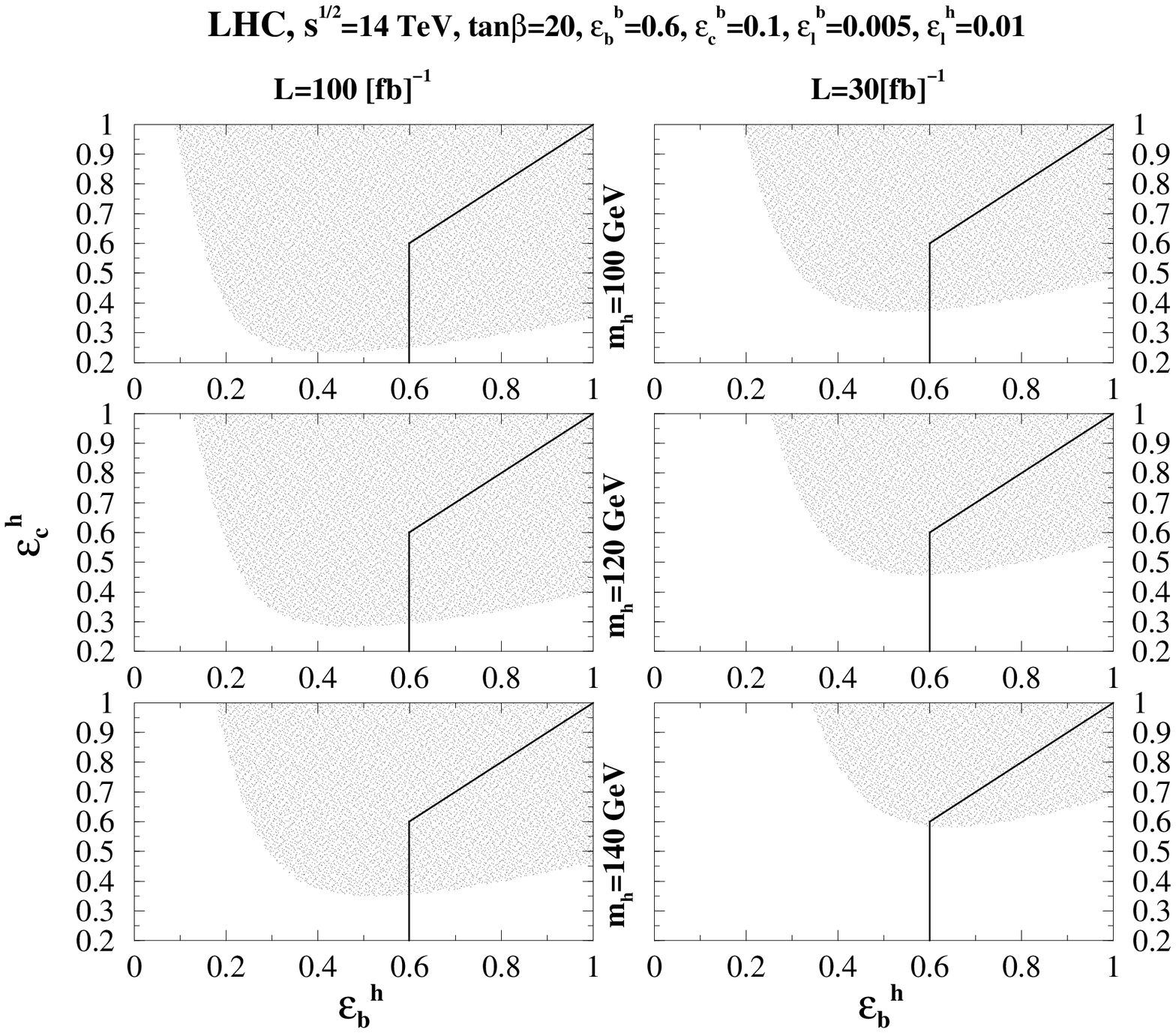,height=17cm,width=15cm,bbllx=0cm,bblly=0cm,bburx=20cm,bbury=20cm,angle=0}
 \end{center}
\caption{Scatter plot for the LHC ($\sqrt{s}=14$ TeV) 
in the $\epsilon_b^h - \epsilon_c^h$ plane, 
for $\tan\beta=20$, 
$\epsilon_l^b=0.005$, $\epsilon_l^h=0.01$, $\epsilon_c^b=0.1$
and $\epsilon_b^b=0.6$. 
The shaded area allows an above 3-sigma detection of $R^{T2HDM}$ at the 
LHC with integrated luminosity of 
30 fb$^{-1}$ (right column) and 100 fb$^{-1}$ (left column) and 
for $m_h=100,~120$ and $140$ GeV on the first, second and third rows, 
respectively. 
The shaded area enclosed to the right of the solid lines corresponds to 
the conditions 
$\epsilon_b^h \geq \epsilon_c^h$ and $\epsilon_b^h \geq \epsilon_b^b$. 
All cross-sections were calculated for 
$\mu=\mu_F=\mu_R=m_h$ and with the set of cuts 1-4 as 
outlined in section \ref{sec3}.}
\label{fig1scatter1}
\end{figure*}

\begin{figure*}[htb]
\psfull
 \begin{center}
  \leavevmode
  \epsfig{file=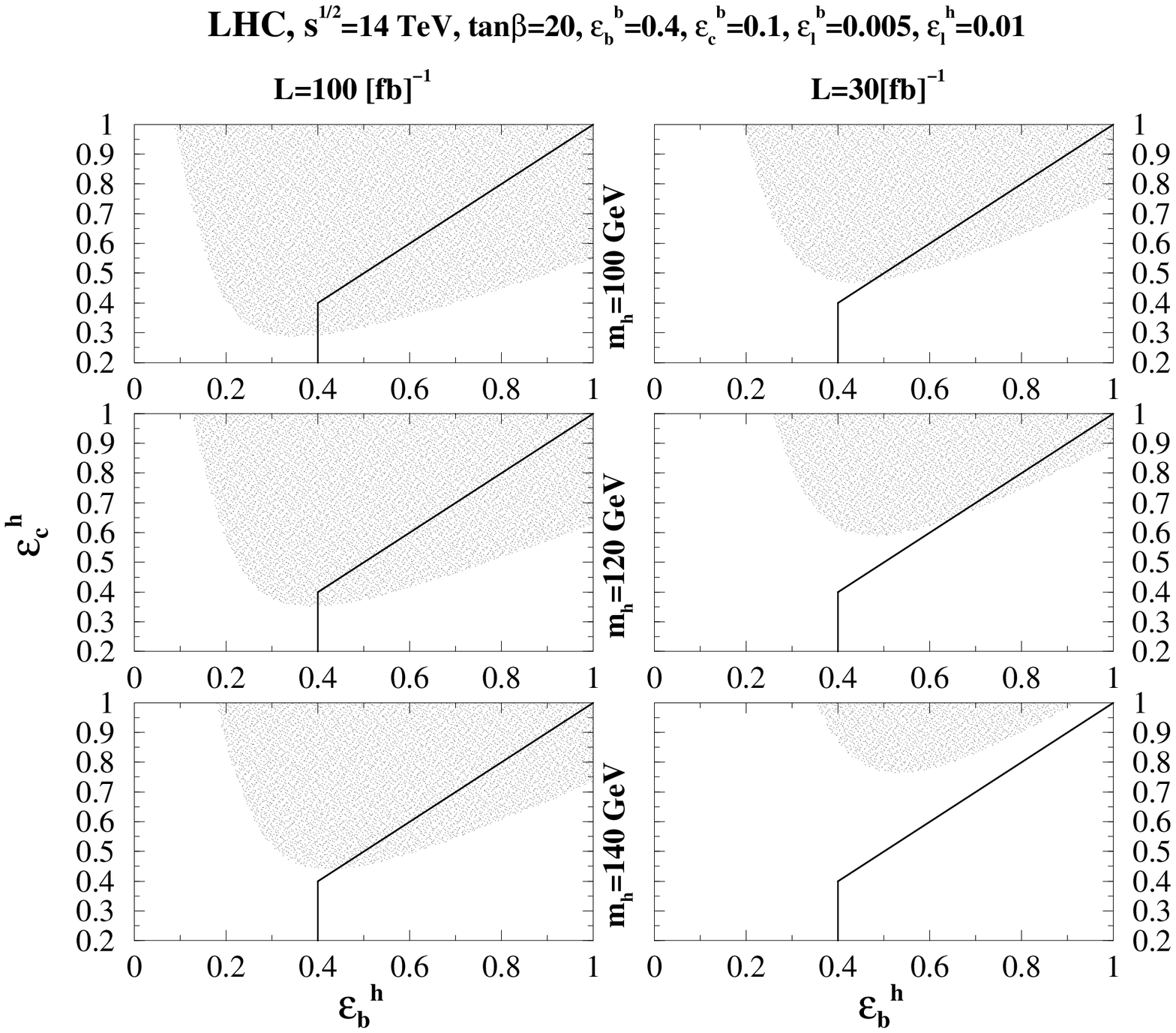,height=17cm,width=15cm,bbllx=0cm,bblly=0cm,bburx=20cm,bbury=20cm,angle=0}
 \end{center}
\caption{Same as Fig.~\ref{fig1scatter1}, for $\epsilon_b^h=0.4$.}
\label{fig1scatter2}
\end{figure*}

Using 
eqs. (\ref{eq6})-(\ref{eq8}) and (\ref{delRexp}), (\ref{NSD}), we plot 
in Figs.~\ref{fig1tb} and \ref{fig2tb} 
the statistical significance ($N_{SD}$) of
$\delta R^{T2HDM}$ (the signal within the T2HDM), as a function of 
$\tan\beta$ for the LHC with an integrated 
luminosity of 30 fb$^{-1}$ and 100 fb$^{-1}$ and for the Tevatron with 
an integrated luminosity of 30 fb$^{-1}$.   
The statistical significance was calculated 
for $\tan\beta=10,~20,~30,~40,~50$ and interpolation was employed to obtain 
the results for other values of $\tan\beta$. 
As our nominal values, 
the following efficiency factors were chosen \cite{epsnumbers}: 
$\epsilon_l^b=0.005$,
$\epsilon_l^h=0.01$,
$\epsilon_c^b=0.1$,
$\epsilon_b^b=0.6$,
$\epsilon_b^h=0.8$, with $\epsilon_c^h=0.6$ in Fig.~\ref{fig1tb} 
and $\epsilon_c^h=0.3$ in  Fig.~\ref{fig2tb}.

Evidently, $\delta R^{T2HDM}$ can be 
easily observed at the LHC for $m_h=100-140$ GeV, 
with a significance of more than 3-sigma
for $\tan\beta \sim {\cal O}(20)$ if $\epsilon_c^h=0.6$ 
and for $\tan\beta \sim {\cal O}(35)$ if $\epsilon_c^h=0.3$.
For example, already at the low luminosity stage of the LHC, i.e., 
with $L=30$ fb$^{-1}$, a 5-sigma detection of 
$\delta R^{T2HDM}$ will be obtained for $m_h \sim 100$ GeV if 
$\epsilon_c^h=0.6$ and $\tan\beta \sim 20$ or 
if $\epsilon_c^h=0.3$ and $\tan\beta \sim 35$. 
On the other hand, a 3-sigma detection of 
$\delta R^{T2HDM}$ will be possible at the Tevatron 
only if $\epsilon_c^h \gsim 0.5$ and if $m_h$ is around 100 GeV and 
$\tan\beta \gsim 50$.
The maximal signal case [obtained when   
$\bar m_b(m_h)$ and $\bar m_c(m_h)$ are evaluated with maximal 
$\bar m_b(\bar m_b)$ and $\bar m_c(\bar m_c)$ as initial conditions, see 
the discussion in section \ref{sec3}], is expected to give a  
3-sigma effect at the Tevatron for $m_h=100$ GeV and $\tan\beta \gsim 45$ 
and if $\epsilon_c^h \sim 0.6$.

In Figs.~\ref{fig1scatter1} and \ref{fig1scatter2} we give 
scatter plots which show the 
region in the $\epsilon_b^h - \epsilon_c^h$ plane that will allow 
a 3-sigma detection of $\delta R^{T2HDM}$ at the LHC 
when $\tan\beta=20$, $m_h=100-140$ GeV,
$\epsilon_l^b=0.005$,
$\epsilon_l^h=0.01$,
$\epsilon_c^b=0.1$,
and with a $b$ jet tagging efficiency of 
$\epsilon_b^b=0.6$ in Fig.~\ref{fig1scatter1} or
of $\epsilon_b^b=0.4$ in Fig.~\ref{fig1scatter2}.
Clearly, a 3-sigma detection of $\delta R^{T2HDM}$ over the 
entire $m_h$ mass 
range shown, 
will be possible at the LHC for a b-tagging efficiency of $60\%$ 
($\epsilon_b^b=0.6$) \cite{epsnumbers} 
and with $\epsilon_b^h,~\epsilon_c^h$ 
as low as $\sim 0.3 - 0.4$ if $L=100$ fb$^{-1}$, 
or with $\epsilon_b^h,~\epsilon_c^h \sim 0.5-0.6$ 
if $L=30$ fb$^{-1}$. 
Fig.~\ref{fig1scatter2} shows that for a lower b tagging 
efficiency 
of $\epsilon_b^b=0.4$, the requirements of         
$\epsilon_b^h, ~ \epsilon_c^h \gsim 0.4$ for $L=100$ fb$^{-1}$ 
and $\epsilon_b^h, ~ \epsilon_c^h \gsim 0.6$ for $L=30$ fb$^{-1}$, will 
also suffice for a 3-sigma detection.

\begin{table}[htb]
\begin{center}
\caption[first entry]{Comparison of the statistical significance 
in the T2HDM ($N_{SD}=\delta R^{T2HDM}/\delta R_{exp}^{T2HDM}$) 
for three values of the factorization scale $\mu_F=m_h/4,~m_h$ 	and $2m_h$.
$N_{SD}$ is given  
for $m_h=100,~120$ and $140$ GeV and $\tan\beta=30$,
at the LHC with an integrated luminosity of 
100 fb$^{-1}$.

\bigskip
\protect\label{tabNSD}}
\begin{tabular}{|c||c|c|c|}
\multicolumn{4}{|c|}{$N_{SD}$ in the T2HDM with $\tan\beta=30$} \\ \hline
& \multicolumn{3}{c|}{LHC with ${\cal L}=100$ fb$^{-1}$} \\ \cline{2-4}
[GeV]$\Rightarrow$& $m_h=100$ & $m_h=120$  &$m_h=140$\\ \hline \hline
$\mu_F=m_h/4$ & $6.37$ & $5.15$ & $4.03$ \\ 
&&&\\
$\mu_F=m_h$ & $6.71$  & $5.14$   & $3.92$ \\ 
&&&\\
$\mu_F=2m_h$ & $6.51$  & $5.03$   & $3.77$ \\ 
&&&\\ \cline{1-4}
\end{tabular}
\end{center}
\end{table}

Finally, in Table \ref{tabNSD} we show the variation of the 
statistical significance of the signal ($N_{SD}$ of
$\delta R^{T2HDM}$) for different choices of the factorization scale, 
$\mu_F=m_h/4,~m_h$ and $2m_h$. The numbers in Table \ref{tabNSD} correspond 
to the LHC with an integrated luminosity of 100 fb$^{-1}$ and are given for
$\tan\beta=30$. Evidently, only mild changes (smaller than 5\%) 
in $N_{SD}$ are obtained when varying the factorization scale 
within the range $m_h/4 \leq \mu_F \leq 2m_h$.  

\subsection{The 2HDMII and the MSSM} 
 
Let us now examine the expected signals 
$\delta R^{2HDMII}$ and  $\delta R^{MSSM}$ in the 2HDMII 
and in the MSSM, respectively.
Here too, we assume from the outset 
that $\tan\beta$ is large enough so that the 
non-standard Higgs can be observed in association with bottom quark 
jets at the hadron
colliders under investigation (recall that if the neutral 
Higgs does not have the necessary $\tan\beta$ 
enhancement in its Yukawa coupling to the bottom quark, then our signal 
ceases to be effective). Thus, in what follows we will focus only on the large 
$\tan\beta$ case. 

The 2HDMII Yukawa couplings follow from the Yukawa potential in 
(\ref{yukawa1}) by setting $U^1=0$ and $D^2=0$. Thus, in this model $\Phi_2$ 
is responsible for the mass generation of the 
up-type quarks while the down-type quark masses are generated through the 
VEV of $\Phi_1$. Denoting again the three neutral Higgs species of the 2HDMII
by $h$, $H$ (the two CP-even scalars) and $A$ (the CP-odd Higgs), their 
couplings to fermions are given by (see e.g., \cite{HHG}):   

\begin{eqnarray}
hd \bar d&:&~Y^{SM}_d \times \left[ \sin(\beta-\alpha)-\tan\beta \cos(\beta-\alpha) \right]~, \nonumber \\ 
hu \bar u&:&~Y^{SM}_u \times \left[ \sin(\beta-\alpha)+\cot\beta \cos(\beta-\alpha) \right]~, \nonumber \\
Hd \bar d&:&~Y^{SM}_d \times \left[ \cos(\beta-\alpha)+\tan\beta \sin(\beta-\alpha) \right]~, \nonumber \\
Hu \bar u&:&~Y^{SM}_u \times \left[ \cos(\beta-\alpha)-\cot\beta \sin(\beta-\alpha) \right]~, \nonumber \\
Ad \bar d&:&~Y^{SM}_d \times \tan\beta \gamma_5 ~, \nonumber \\
Au \bar u&:&~Y^{SM}_u \times \cot\beta \gamma_5 \label{yuk2HDMII}~,
\end{eqnarray}

\noindent where $d$($u$) stands for a down(up) quark and 
$\alpha$ is the mixing angle in the CP-even Higgs sector. 
We will study three representative cases: 

\begin{description}
 
\item{Case 1:} $\cos(\beta-\alpha) \to 1$ or $\alpha \to \beta$.

\item{Case 2:} $\cos(\beta-\alpha) \to 0$ or $\alpha \to \beta -\pi/2$.

\item{Case 3:} $\cos(\beta-\alpha) \sim \sin(\beta-\alpha) \to 1/\sqrt{2}$ or $\alpha \to \beta -\pi/4$.

\end{description}

The charm and bottom Yukawa couplings and their ratio, $Y_c/Y_b$, 
are given in Table \ref{tabratio}, for the three cases above and for each of
the neutral Higgs bosons $h,~H$ and $A$. 

\begin{table}[htb]
\caption[first entry]{The charm and bottom Yukawa couplings $Y_c$ and $Y_b$ 
scaled by their SM Yukawa coupling $Y_c^{SM}$ and $Y_b^{SM}$, respectively, 
and their ratio $|Y_c/Y_b|$, 
in the 2HDM of type II. 
The Yukawa couplings of $h,~H$ and $A$ are given for the 
three cases: (1) $\cos(\beta-\alpha) \to 1$, 
(2) $\cos(\beta-\alpha) \to 0$ and 
(3) $\cos(\beta-\alpha) \sim \sin(\beta-\alpha) \sim 1/\sqrt{2}$, 
see also text.
\bigskip
\protect\label{tabratio}}
\begin{tabular}{|c|c||c|c|c|} 
\multicolumn{5}{|c|}{$Y_c$ and $Y_b$ in the 2HDMII} \\ \hline
Case & Higgs   &  & &  \\
& particle& $|Y_c/Y^{SM}_c|$ & $|Y_b/Y^{SM}_b|$&  $|Y_c/Y_b|$ \\ \hline \hline
& $h$ & $\frac{1}{\tan\beta}$ & $\tan\beta$ & $\frac{m_c}{m_b \tan^2\beta}$ \\ 
1 &$H$ & $1$ & $1$  & $\frac{m_c}{m_b}$ \\ 
& $A$ & $\frac{1}{\tan\beta}$  & $\tan\beta$   & $\frac{m_c}{m_b\tan^2\beta}$ \\ \hline \hline
& $h$ & $1$ & $1$ & $\frac{m_c}{m_b}$ \\ 
2 &$H$ & $\frac{1}{\tan\beta}$ & $\tan\beta$  & $\frac{m_c}{m_b\tan^2\beta}$ \\ 
& $A$ & $\frac{1}{\tan\beta}$  & $\tan\beta$   & $\frac{m_c}{m_b\tan^2\beta}$ \\ \hline \hline
& $h$ & $\frac{1}{\sqrt{2}}$ & $\frac{\tan\beta}{\sqrt{2}}$ & $\frac{m_c}{m_b \tan\beta}$ \\ 
3 &$H$ & $\frac{1}{\sqrt{2}}$ & $\frac{\tan\beta}{\sqrt{2}}$  & $\frac{m_c}{m_b\tan\beta}$ \\ 
& $A$ & $\frac{1}{\tan\beta}$  & $\tan\beta$   & $\frac{m_c}{m_b\tan^2\beta}$ \\ \hline \cline{1-5}
\end{tabular}
\end{table}

The expected signals and their statistical significance for the three limiting 
cases  and for the three neutral Higgs particles of the 2HDMII 
are compared to the T2HDM case in Table \ref{tab2HDMII}.
Clearly, even in the more favorable cases in which 
$N_{SD}(2HDMII) \sim N_{SD}(T2HDM)/\tan^2\beta$, the expected statistical 
significance in the 2HDMII case 
is still at least two orders of magnitudes smaller than the one 
expected in the T2HDM when $\tan\beta \gsim 10$. 
Therefore, based on the results obtained for the T2HDM scenario, we   
conclude that, no signal of $R$ is expected to be observed 
at the LHC or at the Tevatron 
if the Yukawa Higgs sector is controlled by the 2HDMII. 
Hence, a measured signal of $\delta R$ at these hadron colliders 
will {\it rule out} the 2HDMII Higgs scenario with many standard deviations.

\begin{table}[htb]
\caption[first entry]{The expected 2HDMII signals, $\delta R^{2HDMII}$, 
and their 
statistical significance, $N_{SD}^{2HDMII}$, scaled by the 
corresponding values 
in the T2HDM, for cases (1) $\cos(\beta-\alpha) \to 1$, 
(2) $\cos(\beta-\alpha) \to 0$ and 
(3) $\cos(\beta-\alpha) \sim \sin(\beta-\alpha) \sim 1/\sqrt{2}$, 
and for the three neutral Higgs of the model $h$, $H$ and $A$.   
\bigskip
\protect\label{tab2HDMII}}
\begin{tabular}{|c|c||c|c|c|} 
\multicolumn{5}{|c|}{Expected signals in the 2HDMII} \\ \hline
Case & Higgs   &  & &  \\
& particle& $\frac{\delta R^{2HDMII}}{\delta R^{T2HDM}}$ & 
$\frac{\delta R^{2HDMII}_{\rm exp}}{\delta R^{T2HDM}_{\rm exp}}$ & 
$\frac{N_{SD}^{2HDMII}}{N_{SD}^{T2HDM}}$ \\ \hline \hline
& $h$ & $\frac{1}{\tan^4\beta}$  & $1$   & $\frac{1}{\tan^4\beta}$ \\ 
1 &$H$ & $1$ & $\tan^2\beta$  &  $\frac{1}{\tan^2\beta}$ \\ 
& $A$ & $\frac{1}{\tan^4\beta}$  & $1$   & $\frac{1}{\tan^4\beta}$ \\ \hline \hline
& $h$ &  $1$ & $\tan^2\beta$  &  $\frac{1}{\tan^2\beta}$ \\ 
2 &$H$ & $\frac{1}{\tan^4\beta}$  & $1$   & $\frac{1}{\tan^4\beta}$ \\ 
& $A$ & $\frac{1}{\tan^4\beta}$  & $1$   & $\frac{1}{\tan^4\beta}$ \\ \hline \hline
& $h$ & $\frac{1}{\tan^2\beta}$ & $1$ & $\frac{1}{\tan^2\beta}$ \\ 
3 &$H$ & $\frac{1}{\tan^2\beta}$ & $1$ & $\frac{1}{\tan^2\beta}$ \\ 
& $A$ &  $\frac{1}{\tan^4\beta}$  & $1$   & $\frac{1}{\tan^4\beta}$ \\ \hline \cline{1-5}
\end{tabular}
\end{table}

The 2HDMII Higgs framework also underlies the MSSM Higgs sector. 
However, due to the supersymmetric structure of the theory, 
the MSSM's Higgs sector is completely determined at tree-level 
by only two free parameters, conventionally chosen to be $m_A$ and $\tan\beta$. 
That is, the mixing angle $\alpha$ is fixed 
by $m_A$ and $\tan\beta$ at tree level. Concentrating again on the large 
$\tan\beta$ case, in the MSSM one can distinguish two limiting cases:

\begin{description}
 
\item{Case 1:} $m_A \lsim m_h^{max}$, where 
$m_h^{max} \sim 120 ~{\rm or}~135$ GeV, is 
the maximal allowed mass of the lighter CP-even Higgs 
after radiative corrections are included 
in the CP-even sector, depending whether one takes the minimal 
or the maximal stop mixing scenario, respectively \cite{susyhiggs}. 
In this case $\cos(\beta-\alpha) \to 1$.

\item{Case 2:} $m_A^2 >> m_Z^2$, the so called decoupling limit.
In this limit $\cos(\beta-\alpha) \to 0$.

\end{description}  

These two MSSM limiting cases for large $\tan\beta$ 
remain valid also after 
the radiative corrections to the Higgs sector are 
included \cite{susyrad1}, thereby shifting the value of the mixing angle 
$\alpha$ (the higher order contribution to $\alpha$ can, on the other hand, 
cause a large shift 
to the lighter CP-even Higgs mass, $m_h$).$^{[4]}$\footnotetext[4]{There 
are also vertex 
corrections to the Higgs-fermion Yukawa couplings which do not depend
on the mixing angle $\alpha$. However, these have a negligible effect in the
up quark sector, and even more so for large $\tan\beta$ 
values \cite{susyrad2}.}
It should also be noted that, in the large $\tan\beta$ case, 
$\cos^2(\beta-\alpha)$ approaches zero very rapidly as $m_A$ is increased. 
In particular, an order of magnitude  
drop of $\cos^2(\beta-\alpha)$ (from 1 to 0.1) is spanned over no more than  
about $10$ GeV mass range of $m_A$ \cite{susyrad1}.
 
Since cases 1 and 2 in the MSSM Higgs sector are equivalent to 
cases 1 and 2 of the 2HDMII framework, they have the same Yukawa 
couplings pattern as given in Table \ref{tabratio}. Thus, applying the 
results obtained in the 2HDMII to the MSSM case, we conclude
here too that 
$\delta R^{MSSM}$ {\it cannot} be observed either at the 
Tevatron or at the LHC. 
Reversing the argument, a
measured signal of $\delta R$ at these hadron colliders will  
{\it rule out} the MSSM.

\section{Summary}
  
To summarize, we have proposed a signal, 
$\delta R$ [see (\ref{delRth})], based on 
counting the number 
of three heavy ($c$ or $b$) jets  events versus 
the number of events with three $b$ jets, in processes 
in which the neutral Higgs is produced in association with a single high 
$p_T$ $c$ or $b$ jet. This signal assumes that the Higgs Yukawa coupling to 
the $b$ quark is enhanced, say by a large $\tan\beta$ in multi Higgs 
doublet models, and that the neutral Higgs will therefore be 
observed in association
with $b$ jets at future hadron collider experiments.      

This signal was calculated in the framework of multi Higgs models which 
have at least one neutral Higgs with an enhanced Yukawa coupling to the $b$ 
quark when $\tan\beta$ is large.
We have found that in such cases, the signal to background cross-section ratio 
is typically (after applying some useful kinematical cuts) 
$S/B \sim 0.1 - 1$ at the LHC and $S/B \sim 0.05-0.3$ at the Tevatron, 
depending on the value of $\tan\beta$.    
 
The measurement of such a signal will require an efficiency 
for distinguishing a $c$ jet from a light jet at the level 
of about $\epsilon_c^h \sim 20\% -30\%$ or $\epsilon_c^h \sim 50\%-60\%$ at the LHC or at the Tevatron, 
respectively. Such efficiencies seem to be somewhat higher 
than what has been attained in the simulations to date 
($\epsilon_c^h \sim 10\%$)  
\cite{frank}.
We have shown that if such $c$ jet tagging efficiencies are acomplished, 
then our 
signal will be very efficient for probing the 
ratio between the charm and bottom Yukawa couplings, $Y_c/Y_b$, thus allowing 
a deeper insight into the Higgs Yukawa potential. This, in turn, 
will be  useful for categorizing the theory that underlies the Higgs sector. 
It should be noted, that our predictions for the Tevatron rely 
on an accumulated yearly luminosity of 30 inverse [fb], which is, at 
present, considerably higher than the expected Fermilab's run II 
luminosity.    

It should also be noted that the background estimate made for our signal
was based on a low probability for misidentifying a light jet as a
heavy jet, i.e., $\epsilon_l^h \sim {\cal O}(1\%)$. 
While that value is comparable to the present
estimate for $\epsilon_l^b$ (or even a little conservative), 
it is somewhat optimistic for $\epsilon_l^c$
which is not yet well studied by present detector simulations. A more
realistic value for the distinction between a light jet and a charm
jet, based on the technology of today, would be 
$\epsilon_l^c \sim {\cal O}(10\%)$.
Nontheless, our background estimate (and therefore also our estimate for
the significance of the signal) holds roughly for values of
$\epsilon_l^c$ not larger than about 10\%. 
For $\epsilon_l^c$ at the level of tens of percents there may well be 
other processes, e.g., $gg \to ggg$, that may
alter our signal to background estimate to the level that
the observability of our signal becomes difficult.
Another potential problem to the detection of our signal may 
arise if the
trigger algorithm for
b-jet tagging requires a minimum transverse
momentum cut at the level of 100 GeV and above
and the $p_T$ cut of
30 GeV we used for the LHC and 15 GeV for the Tevatron
proves to be too optimistic. In that case
also, the signal proposed in this paper may be  degraded and
difficult to observe due to the large QCD background.

If the difficulties mentioned above can be surmounted and such a signal is 
detected at the Tevatron or at the LHC, 
then the popular MSSM and the 2HDMII can be {\it ruled out} with many standard 
deviations, since in these theories the ratio $Y_c/Y_b$ is too small
for our signal to have any measurable consequences. 
On the other hand, we have shown that within a version of a 
2HDM - the ``2HDM for the top'' 
(T2HDM) -  in which the large mass of the top quark 
is naturally accommodated for large $\tan\beta$, 
our signal 
can be easily observed at the LHC within the entire relevant mass range 
of the neutral Higgs if $\tan\beta \gsim 20$ and 
$\epsilon_c^h \sim 20\% -30\%$. In addition, if the 
neutral Higgs of the T2HDM has a 
mass around 100 GeV and $\tan\beta \gsim 50$, 
then our signal may also give an effect with more than 3-sigma significance 
at the Tevatron provided that $\epsilon_c^h \sim 50\%-60\%$.
         
\begin{acknowledgments}
We would like to thank F. Paige, W. Kilgore and S. Dawson 
for their helpful comments.  
G.E. also thanks 
the Technion President Fund.
This work was also supported in part by US DOE Contract Nos.
DE-FG02-94ER40817 (ISU) and DE-AC02-98CH10886 (BNL).
\end{acknowledgments}

\end{document}